# An Accurate and Fast Start-up Scheme for Power System Real-time Emergency Control

Songhao Yang, *student member, IEEE,* Zhiguo Hao, *Member, IEEE,* Baohui Zhang, *Fellow, IEEE,* Masahide Hojo, *Member, IEEE*

***Abstract*—With the development of PMUs in power systems, the response-based real-time emergency control becomes a promising way to prevent power outages when power systems are subjected to large disturbances. The first step in the emergency control is to start up accurately and fast when needed. To this end, this paper proposes a well-qualified start-up scheme for the power system real-time emergency control. Three key technologies are proposed to ensure the effectiveness of the scheme. They are an instability index, a Critical Machines (CMs) identification algorithm and a two-layer Single Machine Infinite Bus (SMIB) equivalence framework. The concave-convex area based instability index shows good accuracy and high reliability, which is used to identify the transient instability of the system. The CMs identification algorithm can track the changes of CMs and form the proper SMIB system at each moment. The new two-layer SMIB equivalence framework, compared with conventional ones, can significantly reduce the communication burden and improve the computation efficiency. The simulations in two test power systems show that the scheme can identify the transient instability accurately and fast to restore the system to stability after the emergency control. Besides, the proposed method is robust to measurement errors, which enhances its practicality.**

***Index Terms*—Concave-convex area, CMs identification, two-layer SMIB equivalence, transient stability, response-based emergency control**

## I. Introduction

MORDEN power systems are operating near their designed stability limits due to economic reasons and operational constraints [1]. When subjected to large disturbances, these power systems suffer a high risk of loss of synchronism and even power outage. In this situation, the real-time emergency control becomes vital to the power system. The primary step of such control is to identify the transient instability and start the control. The emergency control should only be activated if the system is indeed driving to instability. Otherwise, the wrongly initiated emergency controls may deteriorate the transient stability of the system. In addition, the emergency control should be started fast in order to guarantee the effectiveness of control measures. Thus, a start-up scheme that features accurate and fast transient instability detection is necessarily required for the emergency control.

Songhao Yang, Zhiguo Hao and Baohui Zhang are with School of Electrical Engineering, Xi'an Jiaotong University, Xi'an, China.
(email: songhaoyang@stu.xjtu.edu.cn; zhghao@mail.xjtu.edu.cn; bhzhang@mail.xjtu.edu.cn).
Masahide Hojo are with Department of Electrical and Electronic Engineering, Tokushima University, Tokushima, Japan
(email: hojo@ee.tokushima-u.ac.jp)

Currently, the start-up schemes of the emergency control are either event-based or response-based. The event-based emergency control methods first evaluate the transient stability of a series of pre-imagined contingencies by Transient Stability Assessment (TSA) techniques, e.g., the Extended Equal Area Criterion (EEAC) [2], Single Machine Equivalent (SIME) [3], stability region theory [4] and relative transient energy function [5]. Then they design proper control actions for unstable cases and finally derive a decision table [6, 7]. Once a pre-imagined contingency occurs, the emergency control will be initiated immediately, and the corresponding control actions will be taken [8]. Despite the fast response speed, these event-based emergency control schemes have restricted accuracy due to their dependence on offline simulations when they are applied in the real-time emergency control. Besides, these methods can only handle the pre-imagined contingencies, but fail to deal with the unexpected ones.

The response-based real-time emergency control, by contrast, aims at assessing whether the contingency that has already occurred is driving the system to instability. With the development of PMUs in the power system, these response-based methods are attracting more and more attention. These methods can be classified into three categories:1) predicted response base methods, 2) machine learning based methods and 3) instability characteristic based methods.

The predicted response based start-up schemes predict the future transient response of the power system by model-free methods such as the curve fitting technique [9] and then detect the instability in real time. Despite being simple and requiring little information, these schemes have unstable accuracy because the prediction accuracy is sensitive to the number of data points, the sampling frequency and the order of prediction model [10, 11]. The inaccurate prediction will lead to the malfunction of the real-time emergency control.

Although the machine learning methods are heuristic, they are widely used in emergency control because of their fast response speed. In this kind of methods, a series of classifiers are trained offline through machine learning techniques such as decision tree [12, 13], artificial neural network [1, 14-16] and support vector machine methods [17-20]. Then the classifiers are applied in real time to identify the transient instability. The accuracy of these methods is promising when classifiers are sufficiently trained. However, offline training is a time-consuming work because it needs to adapt to the various system operating conditions and numerous possible disturbances. Moreover, the classifiers become unreliable if the practical system shows different dynamics from the offline training data.

For the instability characteristic based methods, specific instability criteria deduced from the mathematical or physical analysis are involved. They seem to be more reliable than those heuristic methods because of their physical theoretical foundation. The instability criteria are mostly calculated based on the measured physical variables such as the state variables of the buses and generators. In [21], the Lyapunov Exponents (LEs) of generator pairs were applied to identify the transient stability in real time. However, it takes seconds to correctly identify the instability, which is too late for the real-time emergency control. A post-disturbance bus voltage magnitude-based stability prediction method was proposed in [22]. Although the method is fast and straightforward, it needs to determine the stability boundary in advance by offline simulation, where the parameter and model uncertainties may decrease the accuracy. The emergency SIME methods in [3] and [23] get rid of the dependence on models by determining the parameters of the equivalent Singe Machine Infinite Bus (SMIB) system with PMU data. However, the accuracy of the parameters identification cannot be guaranteed in the practical power system. The graphical characteristic of the phase trajectory was applied to the transient instability detection of the SMIB system in [24], which was described as:' a stable trajectory is always concave with respect to the post-fault Stable Equilibrium Point (SEP) while an unstable trajectory is convex with respect to the post-fault SEP immediately or a short time after the fault-clearing time'. The instability feature was mathematically demonstrated in [25] and was further extended to the multi-machine power system by a SMIB equivalence in [26].

The instability characteristic of the phase trajectory is simple and easy to identify, which is promising in the real-time TSA. However, several problems need to be addressed before the characteristic is used to start real-time emergency control. Firstly, when using the discrete PMU data for the instability index calculation, the existing methods in [25, 26] have the severe problem of noise due to multiple differential operations, which may start the emergency control incorrectly. Secondly, the SMIB equivalence is required to evaluate the transient stability of the multi-machine power system. However, the time-changing critical machines (CMs) increase the difficulty of obtaining an accurate SMIB system. Thirdly, in the conventional SMIB equivalence methods, all generator data needs to be directly uploaded to the global control center. However, this way of processing data is not allowed in a practical power system due to data privacy and limitations of existing hierarchical control framework. Moreover, when applied to a large-scale power system, the conventional equivalence methods have the problems of heavy communication burden and long computation time [27, 28]. The transmission of massive data under insufficient bandwidth will cause communication delay and data dropout, which may reduce the accuracy of the start-up scheme. And the long processing time of massive data in the global control center will delay the start-up of the emergency control.

In this paper, an accurate and fast start-up scheme is proposed for power system real-time emergency control. The start-up scheme is continuously executed with the updated measurement data from PMUs. If the power system is identified as unstable, the emergency control will be initiated immediately to prevent the system from collapse. The accuracy of the scheme is guaranteed by a reliable instability index and a real-time CMs identification algorithm. The instability index features less information required, smaller computation amount and higher noise tolerance. The CMs identification algorithm can track the changes of CMs during the transient process to form the correct SIMB system at each moment. The real-time operation of the start-up scheme benefits from a novel two-layer SMIB equivalence framework which enables to reduce the communication burden and improve the computation efficiency of massive data. The rest of paper is organized as follows: Section II introduces the contributions of the work. Section III presents the outline of the start-up scheme for real-time emergency control. Simulations in section IV verify the effectiveness of the proposed start-up scheme and section V gives the conclusion.

## II. Contributions

To start the real-time emergency control accurately and fast, we propose a reliable instability index, a fast two-layer SMIB equivalence framework and a real-time critical machines identification algorithm in this paper.

### A. *The concave-convex area-based instability index*

For a SMIB power system, the convex-concavity of the phase trajectory is proved to be related to the system's transient stability [24-26]. To apply this relationship to real-time emergency control of the power system, we develop a novel instability index based on the concept of the concave-convex area on the phase plane.

Mathematically, the curve $y = f(x)$ is convex if it satisfies $f''(x) > 0$. For the phase trajectories on the phase plane that is composed of the generator's power angle $\delta$ and angular speed deviation $\Delta\omega$, the concave-convexity index of the phase trajectory can be defined as

$$l = {d^2\Delta\omega}/{d\delta^2} \quad (1).$$

The continuous curve composed by the phase points that satisfy $l$=0 is called inflection curve. It naturally divides the phase plane into two areas: the concave area $(l \bullet \Delta\omega < 0)$ and the convex area $(l \bullet \Delta\omega > 0)$. The location of the concave-convex areas on the phase plane is given in Fig. 1 which also shows the relationship between phase trajectories and the concave-convex areas. In Fig. 1, the trajectories of stable cases in which the fault duration is shorter than the Critical Clearance Time (CCT) ($t_{c1}<t_{c2}<CCT$) always stay in the concave area whereas the trajectories of unstable cases in which the fault is cleared later than the CCT ($CCT<t_{c3}<t_{c4}$) enter the convex area inevitably. As a 'black hole' on the phase plane, the convex area is the no-return-area for phase trajectories, which indicates the transient instability of the power system. Thus, transient instability of the system can be identified by detecting whether the phase trajectory enters the convex area.

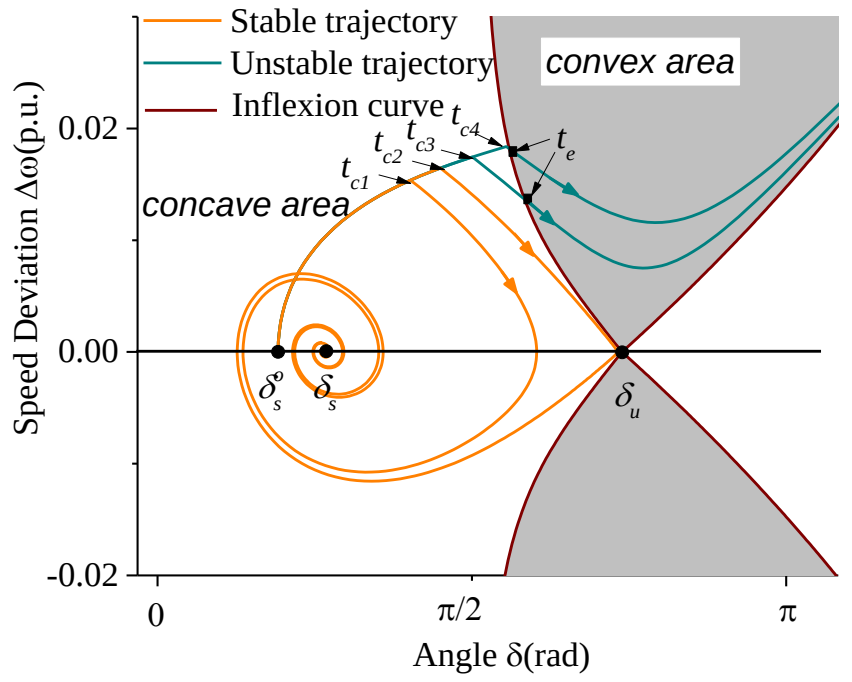

Fig. 1 Phase trajectories and the concave-convex areas on the phase plane

According to the definition of the convex area, the phase trajectories in convex area satisfy (2).

$$l \bullet \Delta\omega > 0 \Rightarrow \frac{d^2\Delta\omega}{d\delta^2} \bullet \Delta\omega > 0 \Rightarrow \frac{dk}{d\delta} \bullet \frac{d\delta}{dt} \bullet \frac{1}{\omega_0} > 0 \Rightarrow \frac{dk}{dt} > 0, \quad (2)$$

where $k$ is the slope of phase trajectory $k = {d\Delta\omega}/{d\delta}$ and $\omega_0 = 2\pi f_0$. $f_0$ is the system frequency, 50Hz or 60Hz.

The rotor motion equations of the SMIB system in (3) is used to avoid the multi-order derivation operations in (2).

$$\dot{\delta} = \omega_0 \Delta\omega \,;\, M\,\dot{\Delta\omega} = P_m - P_e - D\omega_0\Delta\omega = \Delta P - D\omega_0\Delta\omega, \quad (3)$$

where $P_e$, $P_m$ and $\Delta P$ are separately generator's electrical power, mechanical power, and power difference. Parameters $D$ is the damping factor, and $M$ is inertia constant.

Thus, the slope $k$ can be written in the form of (4).

$$k = \frac{d\Delta\omega}{dt} \bullet \frac{dt}{d\delta} = \frac{\Delta P - D\omega_0\Delta\omega}{M\omega_0\Delta\omega} \quad (4)$$

Since that the measurement data from PMUs is discrete, (5) gives a differential form of the new instability criterion in (2).

$$\frac{k(t) - k(t-\Delta t)}{\Delta t} = \frac{1}{\Delta t M \omega_0}\left(\frac{\Delta P(t)}{\Delta\omega(t)} - \frac{\Delta P(t-\Delta t)}{\Delta\omega(t-\Delta t)}\right) > 0 \quad (5)$$

A simpler form is deduced in (6) by omitting the constant terms in (5).

$$c(t) = \frac{\Delta P(t)}{\Delta\omega(t)} - \frac{\Delta P(t-\Delta t)}{\Delta\omega(t-\Delta t)} > 0 \quad (6)$$

It can be inferred from (6) that the calculation of the new instability index only require the data of the current moment $t$ and last moment $t-\Delta t$, and all required generator information, $\Delta P$ and $\Delta\omega$, can be obtained from PMUs.

Also, it should be noted in Fig. 1 that the instability detection time, $t_e$, namely the moment phase trajectory enters the convex area, is earlier than the moment that the angle exceeds the angle of Unstable Equilibrium Point(UEP), $\delta_u$. According to [29], if no control measures are taken, the system will inevitably lose the stability after the phase angle exceeds $\delta_u$. In other words, the proposed instability criterion can predict transient instability before the system becomes truly unstable. The early identification of instability leaves enough time to initiate and implement the emergency control.

### *B. The two-layer SMIB equivalence framework*

The SMIB equivalence is widely used in TSA to identify the transient stability of the multi-machine power system [3]. It assumes that the loss of synchronism originates from the separation of generators into two groups: the group $C$ of critical machines (CMs) and the other group $N$ of non-critical machines (NMs). The transient stability of the original multi-machine power system is presented by the relative motion of these two groups of generators. Therefore, the equivalent SMIB system describing the relative motion in (7) is necessary for the TSA.

$$\delta_r = \frac{\sum_{i\in C} M_i\delta_i}{\sum_{i\in C} M_i} - \frac{\sum_{j\in N} M_j\delta_j}{\sum_{j\in N} M_j}, \Delta\omega_r = \frac{\sum_{i\in C} M_i\Delta\omega_i}{\sum_{i\in C} M_i} - \frac{\sum_{j\in N} M_j\Delta\omega_j}{\sum_{j\in N} M_j},$$
$$M_r = \frac{\sum_{i\in C} M_i \sum_{j\in N} M_j}{\sum_{i\in C} M_i + \sum_{j\in N} M_j}, \Delta P_r = \frac{\sum_{j\in N} M_j \sum_{i\in C}\Delta P_i - \sum_{i\in C} M_i \sum_{j\in N}\Delta P_j}{\sum_{i\in C} M_i + \sum_{j\in N} M_j}, \quad (7)$$

where $\delta, \Delta\omega, M$ and $\Delta P$ are the power angle, angular speed deviation, the inertia constant and the power difference, respectively. The variables attached with the subscript $i$ represent the state variables of the $i$-th generator and those attached with subscript $r$ represent the state variables of the equivalent SMIB system.

To address the problems of heavy communication burden and long computation time in the conventional SMIB equivalence method, we propose a novel two-layer SMIB equivalence framework in this paper.

We assume that the practical power grid consists of several area grids. There is one area control center for each area grid and one global control center for the entire power grid. Note that the concept of area grid is different from that of generator group. Generator location decides its belonging to the area grid whereas the generator dynamic decides its belonging to the generator group (CMs or NMs). The relationships between generators, area grids, and generator groups are shown in Fig. 2. Generators in same area grid may belong to same or different generator groups during the transient process. For example, all generators of area 1 (area 3) belong to the same generator group $C$ (group $N$), whereas generators of area 2 belong to different generator groups.

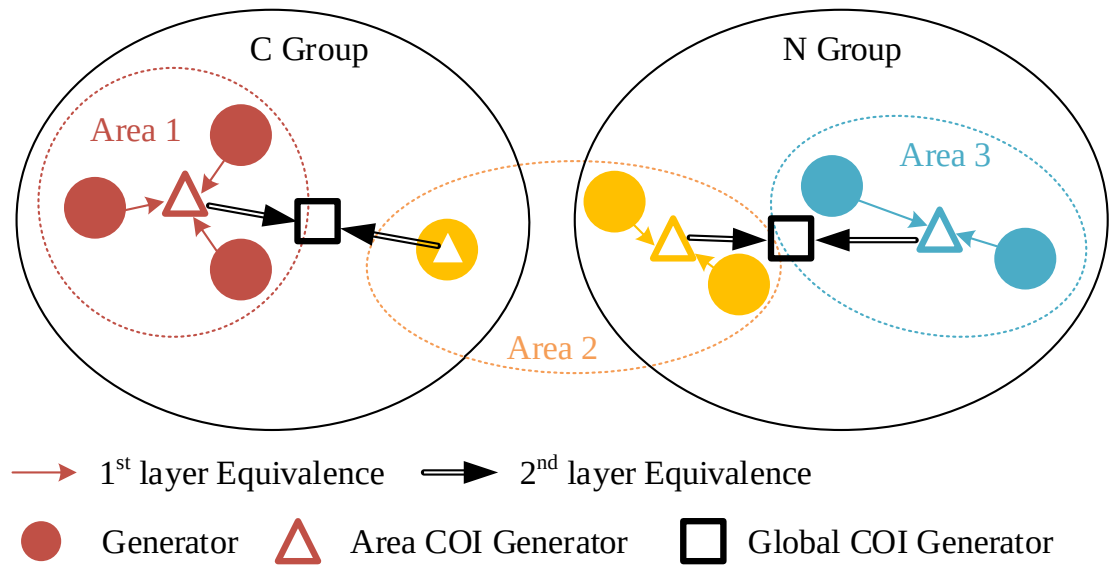

Fig. 2 The chart of the two-layer SMIB equivalence framework

Based on the assumed two-layer control structure, the two-layer SMIB equivalence framework is presented as:

a) 1st layer equivalence: In each area grid, generators that belong to same generator group are equivalent to one Center of Inertia (COI) generator by (8).

$$\delta_G^p = \frac{\sum_{i\in p, i\in G} M_i\delta_i}{\sum_{i\in p, i\in G} M_i}, \Delta\omega_G^p = \frac{\sum_{i\in p, i\in G} M_i\Delta\omega_i}{\sum_{i\in p, i\in G} M_i}$$
$$M_G^p = \sum_{i\in p, i\in G} M_i, \Delta P_G^p = \sum_{i\in p, i\in G} \Delta P_i \quad (8)$$

where $\delta_G^p, \Delta\omega_G^p, M_G^p$ and $\Delta P_G^p$ are separately the power angle, angular speed deviation, the inertia constant and the power difference of the COI generator in area $p$ and group $G$ ($G$=$C$ or $N$). As Fig. 2 shows, generators that belong to the same group, such as those in area 1 or 3, are equivalent to one COI generator. And generators that belong to different groups, such as those in area 2, are equivalent to two COI generators.

b) 2nd layer equivalence: The COI generators that belong to the same generator group($C$ or $N$) are equivalent to one global COI generator. The two global COI generators compose the final equivalent SMIB system by (9).

$$\delta = \frac{\sum_p M_C^p \delta_C^p}{\sum_p M_C^p} - \frac{\sum_p M_N^p \delta_N^p}{\sum_p M_N^p}, \Delta\omega = \frac{\sum_p M_C^p \Delta\omega_C^p}{\sum_p M_C^p} - \frac{\sum_p M_N^p \Delta\omega_N^p}{\sum_p M_N^p},$$
$$M = \frac{\sum_p M_C^p \sum_p M_N^p}{\sum_p M_C^p + \sum_p M_N^p}, \Delta P = \frac{\sum_p M_N^p \sum_p \Delta P_C^p - \sum_p M_C^p \sum_p \Delta P_N^p}{\sum_p M_C^p + \sum_p M_N^p} \quad (9)$$

It can be mathematically proved that the final equivalent system obtained by the two-layer SMIB equivalence method in (8) and (9) is identical to the one obtained by the direct SMIB equivalence method in (7). In other words, the new SMIB equivalence method does not affect the accuracy of transient instability detection. Besides, the two-layer SMIB equivalence method features less communication burden and higher computation efficiency because the area control centers share most of the computation. Further analysis and verification are given in section IV subsection C.

### *C. The real-time CMs identification algorithm*

In the two-layer SMIB equivalence framework, the CMs identification is required in both the area control centers and the global control center. Accurate CMs identification is the guarantee of the correct SMIB equivalence. When applied to the real-time emergency control, the CMs identification encounters stricter requirements. First, the CMs identification algorithm should be fast and straightforward to reduce the computation time. Besides, the identification results should be continuously updated over time to track the changes of CMs. To meet these requirements, we propose a real-time CMs identification algorithm based on the Largest Angle Gap (LAG).

*step:1* Initialize all generators at $t$: $\delta_i'(t) = \delta_i(t) - \delta_i(0)$ ;

*step:2* Sort generators in increasing order by $\delta_i'(t)$ ;

*step:3* Compute the angle gap between nearby generators in the sequence obtained in *step 2*;

*step:4* Choose the LAG as the boundary of different generator groups. Generators above the boundary belong to the $C$-group, and others belong to the $N$-group.

Note that we pay more attention to the post-disturbance behavior of generators. Thus we adopt the initialization in *step 1* to avoid the dependence on the initial state, just like [30].

The primary feature of the proposed algorithm is that the results of CMs identification are continuously updated over time. The feature allows the algorithm to track the changes of CMs and obtain an accurate equivalent SMIB system at each moment. Moreover, the real-time operation of CMs identification also helps to make more effective emergency control strategies. Accurate CMs identification means correct control objects, which are the guarantee of effective emergency control.

## III. The Start-up Scheme For Real-time Emergency Control

In this paper, a start-up scheme for real-time emergency control is proposed based on the two-layer control structure in a power system. The area control centers collect the measurement data from PMUs and transmit the data to the global control center. The global control center is responsible for the transient instability detection and the emergency control start-up. Fig. 3 (a) and (b) show the schematic diagram in the area control centers and the global control center, respectively. Started after the disturbance occurrence, the scheme is executed at each moment with continuously updated measurement data from PMUs. The outline of the start-up scheme is given as follows.

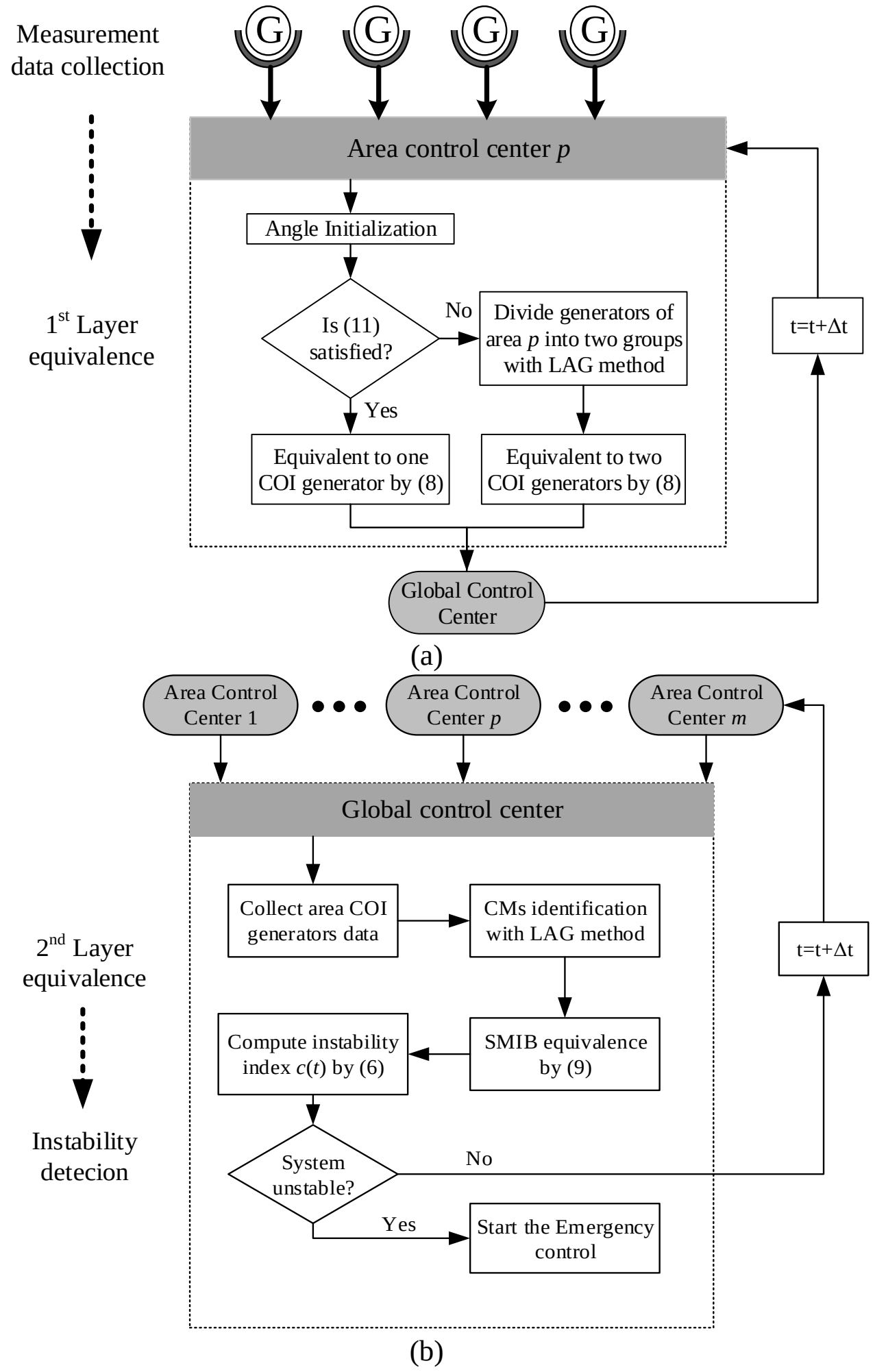

Fig. 3 The schematic diagram in (a) the area control center $p$ and (b) the global control center

### *A. Measurement data collection in the area control centers*

The proposed scheme requires generators' state information including the angle $\delta_i$, angular speed deviation $\Delta\omega_i$ and power difference $\Delta P_i$ at each moment. The PMUs equipped on each generator can provide all the required information. The power angle $\delta_i$ can be obtained directly, and others can be derived indirectly by (10).

$$\Delta\omega(t)=2\pi\frac{f(t)-f_0}{f_0}, \Delta P(t)=P_m(t)-P_e(t) \quad (10)$$

In (10), $f$ is the frequency of the generator bus, and $f_0$ is the system nominal frequency. $P_m$ is the mechanical power which can be measured by the sensors on the rotors. $P_e$ is the electrical power which is provided by PMUs. If it is difficult to measure $P_e$ directly, a backup scheme is given as $P_e = UI\cos\theta$ where $U, I$ and $\theta$ are the voltage, current and the angle between them. The measurement data is updated every recording cycle which is selectable from 20ms to 100ms according to [31]. Then the data will be transmitted to the area control centers for the 1st layer equivalence.

### *B. 1st layer equivalence in the area control centers*

After receiving the measurement data, the area control centers first initializes the angles to reduce the influence of the initial state and then determines whether the generators of this area belong to the same generator group by (11).

$$\delta_{\max,p}(t)-\delta_{\min,p}(t)<\delta_{Set}, \quad (11)$$

where $\delta_{\max,p}(t)$ and $\delta_{\min,p}(t)$ are separately the maximum angle and minimum angle of all generators in the area $p$ at moment $t$. $\delta_{set}$ is the power angle threshold value which is determined by the system scale. The value is suggested in the range of 5°-10°. If (11) is satisfied at moment $t$, all generators of area $p$ are coherent and surely belong to the same generator group [32]. Then these generators are equivalent to one area COI generator by (8). Otherwise, the generators in the area $p$ may belong to two different groups, just like the generators in area 2 in Fig. 2. In such situations, the generators of area $p$ are divided into two groups by the real-time CMs identification algorithm. Then these two generator groups are equivalent to two area COI generators by (8).

By the 1st layer equivalence in the area control centers, generators of each area grid are equivalent to one or two area COI generators. Afterward, the data of area COI generators is transmitted to the global control center for the transient instability identification.

### *C. 2nd layer equivalence in the global control center*

In the global control center, all area COI generators are treated as independent generators, and they are divided into two groups in real time by the LAG based CMs identification algorithm. Then the final equivalent SMIB system at moment $t$ is obtained by (9).

### *D. Transient instability detection in the global control center*

The instability index, $c(t)$, is computed by (6) based on the phase trajectory of the equivalent SMIB system. If $c(t)<0$, the scheme will return to the area control centers and continue to detect the transient instability with updated measurement data of the next moment. If $c(t)>0$, the system is identified as unstable, and the emergency control will be started immediately.

## IV. SIMULATIONS

The start-up scheme for real-time emergency control is tested in the 39-bus 10-machine power system [33] and the 145-bus 50-machine power system [34]. In the 39-bus 10-machine system, all generators use the 4th order generator models, and each of them is equipped with the type ac-4 exciter [35] and the Power System Stabilizer (PSS). In the 145-bus 50-machine system, 6 generators use the 4th order generator models, and the rest 44 generators use the classical generator models. Type ac-4 exciter is also equipped on each generator. The transient simulations are calculated on the PSASP-a software platform for power system simulation. The steady power flow of the power system is first calculated before the transient simulation starts. Then the required data, which includes the power angle, angular speed and power difference of each generator, is simulated in PSASP and inputted as the PMU measurement data into the scheme. The recording rate of the PMUs adopts 50Hz and $\delta_{set}$ in (11) is set as 10°.

### *A. 39 bus 10-machine power system*

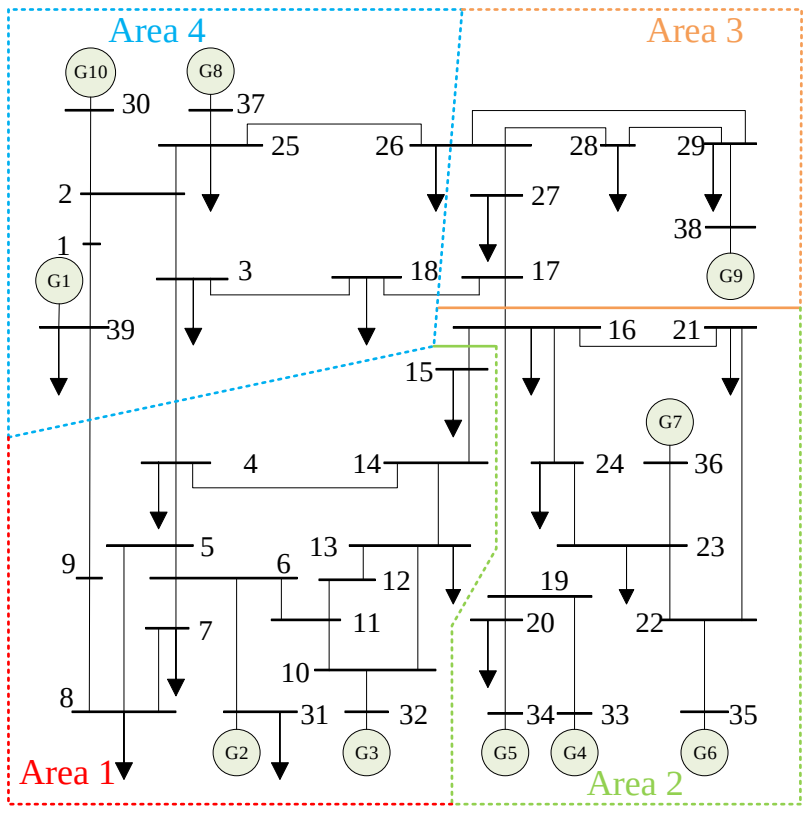

Fig. 4 39-bus 10-machine power system

As Fig. 4 shows, the 39-bus power system consists of 4 areas and 10 generators. Two critical cases are simulated to verify the accuracy of the proposed scheme.

Case 1: Three-phase short-circuit ground fault occurs on the line 26-29 at $t$=0s, then the fault line is eliminated at $t$=0.117s.

Case 2: Same fault occurs on the same line, but the fault line is eliminated at $t$=0.118s. The slight difference in fault duration leads to the opposite transient stability results.

The transient stability of the two cases is identified in real time by the proposed scheme following steps in Fig. 3. After the 1st layer equivalence in the area control centers, the generators of each area grid are equivalent to one or two COI generators, as shown in Fig. 5. It is noted that the angle curves of some area COI generators are segmented because the number of COI generators determined by (11) is not constant during the whole transient process. For example, for the generators in area 4 in case 1, they are equivalent to one COI generator when the power angles are close, but generators are equivalent to two COI generators at moments that (11) is not satisfied. The time-updated 1st layer equivalence ensures that generators belonging to different groups can be separated correctly. It is the guarantee of the accurate CMs identification in the 2nd layer equivalence.

Table 1 shows the results of the real-time CMs identification in the 2nd layer equivalence. The COI Gen. No. is short for the COI generator number. No. 4(5) represents the one (or two) area COI generator(s) in area 4 in case 1 at different moments. Each COI generator is treated as an independent generator in the global control center and then grouped in real time by the CMs identification algorithm. Table 1 shows that CMs change twice and finally form three different instability modes in case 1 whereas the instability mode remains unchanged in case 2.

TABLE 1 RESULTS OF THE REAL-TIME CMS IDENTIFICATION IN CASES 1 AND 2

| | Time | CMs (COI Gen. No.) | NMs (COI Gen. No.) | Mode |
|---|---|---|---|---|
| Case 1 | 0.12s-1.42s | {3}, | {1,2,4(5)} | 1 |
| | 1.44s-1.48s | {1,3,4} | {2} | 2 |
| | 1.50s-2s | {1 2 4(5)}, | {3} | 3 |
| Case 2 | 0.12s-2s | {3} | {1,2,4} | 1 |

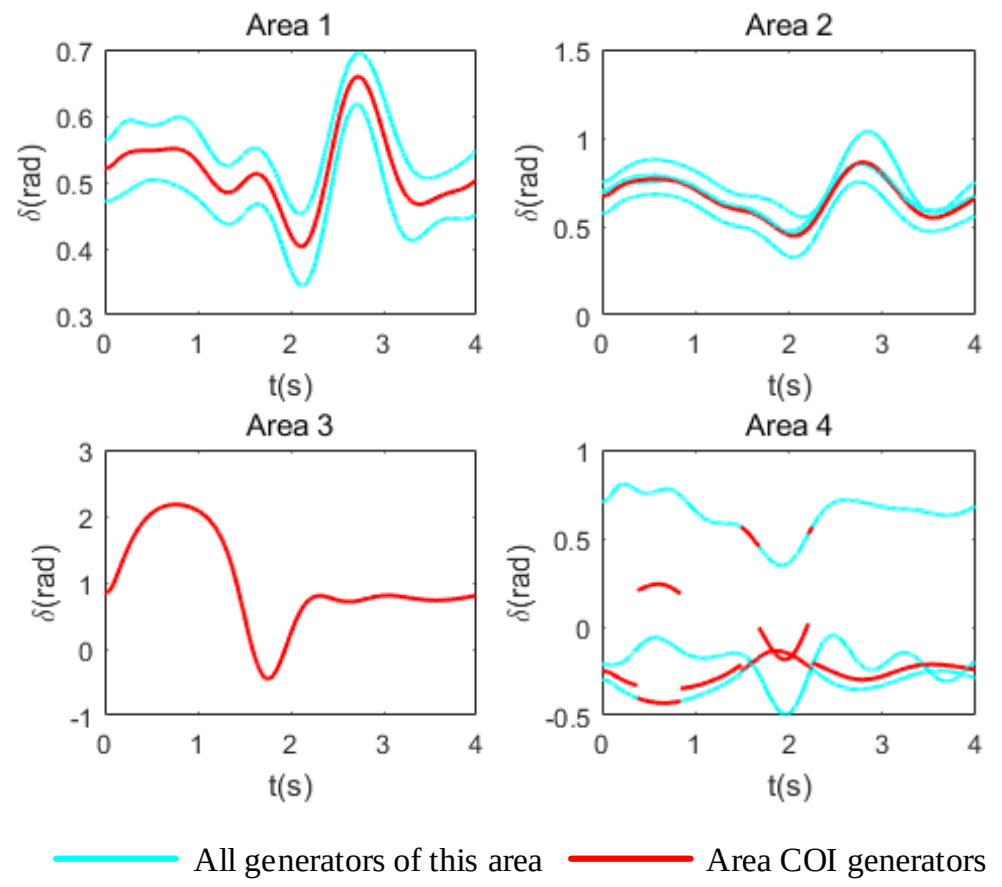

Fig. 5 1st layer equivalence in area control centers in case 1

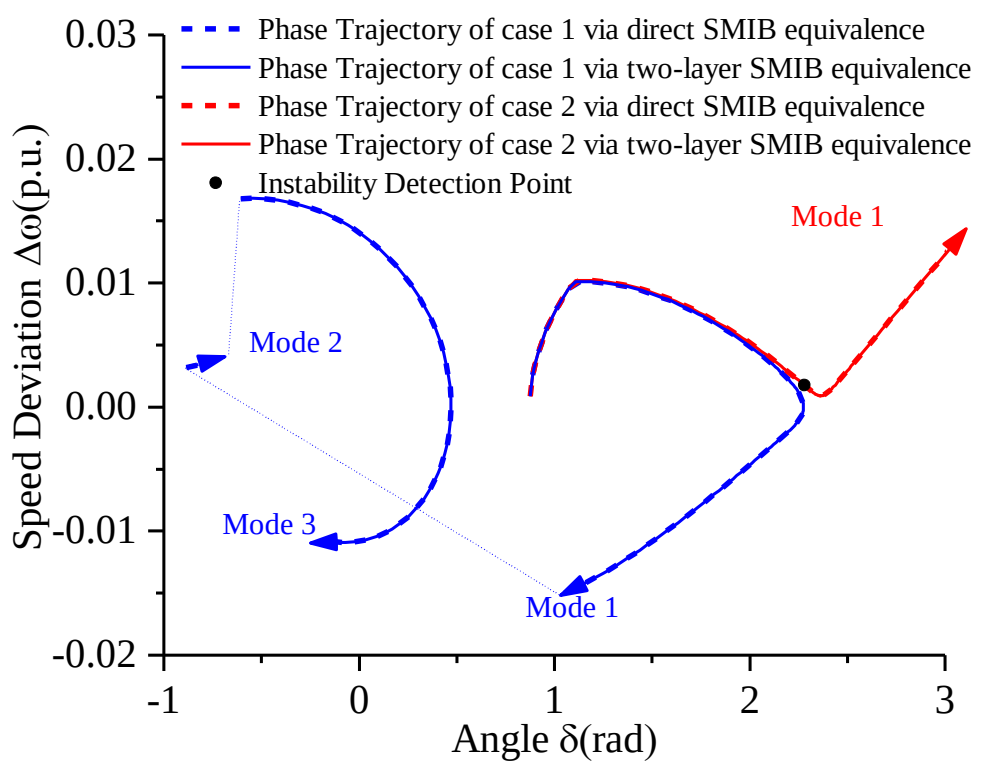

Fig. 6 Phase trajectories of the equivalent SMIB systems in case 1 and 2

Fig. 6 gives the final SMIB phase trajectories after the 2nd layer equivalence. Solid lines show the phase trajectories obtained by the proposed two-layer SMIB equivalence method. By contrast, dash lines show the phase trajectories obtained by direct SMIB equivalence method. As Fig. 6 shows, these phase trajectories obtained by two different ways are entirely identical, which demonstrates the validity of the two-layer SMIB equivalence framework. It is noted that the phase trajectories in case 1 are segmented because of the changes of CMs.

The instability index $c$ is computed at each moment to identify the transient stability in real time, as shown in the blue line in Fig. 7 (a) and (b). It shows that $c$ is always negative in case 1, indicating that case 1 is transient stable. In case 2, $c$ changes from negative to positive at 0.65s, indicating that the transient instability is detected at this moment. As Fig. 6 shows, the phase trajectories in the two critical cases are quite similar in the beginning stage after the disturbance. However, the proposed instability index can still correctly identify the different transient stability in two cases.

To highlight the accuracy and practicality of the proposed instability index, we compare it against other convexity-concavity based instability indexes. The instability indexes involved can refer to [24] and [26]. Table 2 summarizes the formula, required variables, number of data, and instability criterion of three instability indexes. It shows that the index $\mu$ in [24] requires extra information about the post-fault stable equilibrium point which is, however, difficult to obtain in the practical power system. The index $\tau$ in [26] requires the most data, and the calculation includes multiple differential operations. Compared with these two indexes, the index $c$ proposed in this paper requires the least data and takes the minimum calculation.

TABLE 2 COMPARISON OF THREE INSTABILITY INDEXES BASED ON CONVEXITY-CONCAVITY OF PHASE TRAJECTORY

| Instability index | required variables | number of data | instability criterion |
|---|---|---|---|
| $\mu = \dfrac{[-\dot{\omega}(t-\Delta t)\ \omega(t-\Delta t)]\begin{bmatrix}\delta(t-\Delta t)-\delta^{sep} \\ \omega(t-\Delta t)\end{bmatrix}}{[-\dot{\omega}(t-\Delta t)\ \omega(t-\Delta t)]\begin{bmatrix}\delta(t)-\delta^{sep} \\ \omega(t)\end{bmatrix}}$ | $\delta, \omega, \dot{\omega}, \delta_{sep}$ | 2 | $\mu < 1$ |
| $\tau = \dfrac{\Delta\omega(t)-\Delta\omega(t-\Delta t)}{\delta(t)-\delta(t-\Delta t)} - \dfrac{\Delta\omega(t-\Delta t)-\Delta\omega(t-2\Delta t)}{\delta(t-\Delta t)-\delta(t-2\Delta t)}$ | $\delta, \Delta\omega$ | 3 | $\tau > 0$ |
| $c = \dfrac{\Delta P(t)}{\Delta\omega(t)} - \dfrac{\Delta P(t-\Delta t)}{\Delta\omega(t-\Delta t)}$ | $\Delta\omega, \Delta P$ | 2 | $c > 0$ |

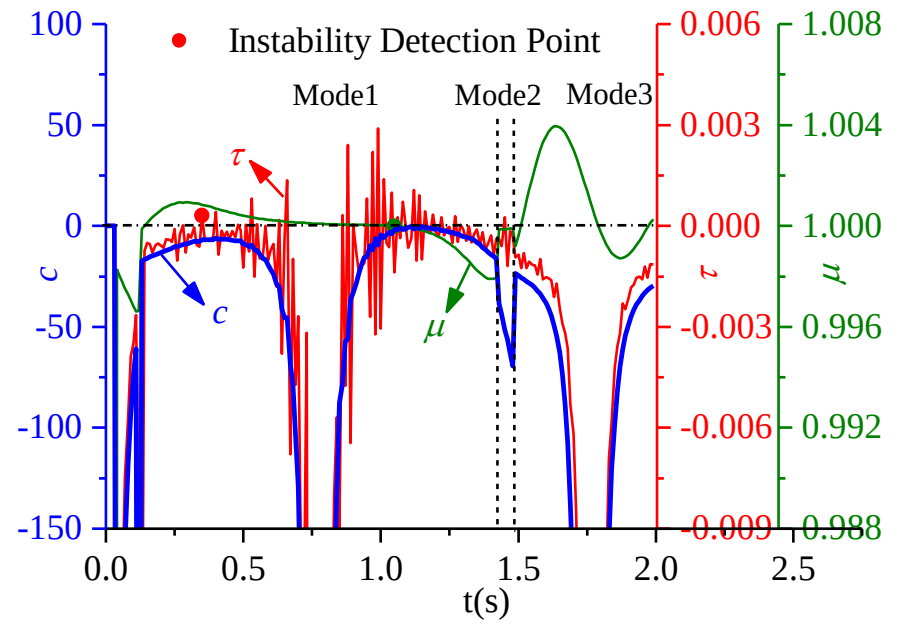

(a)

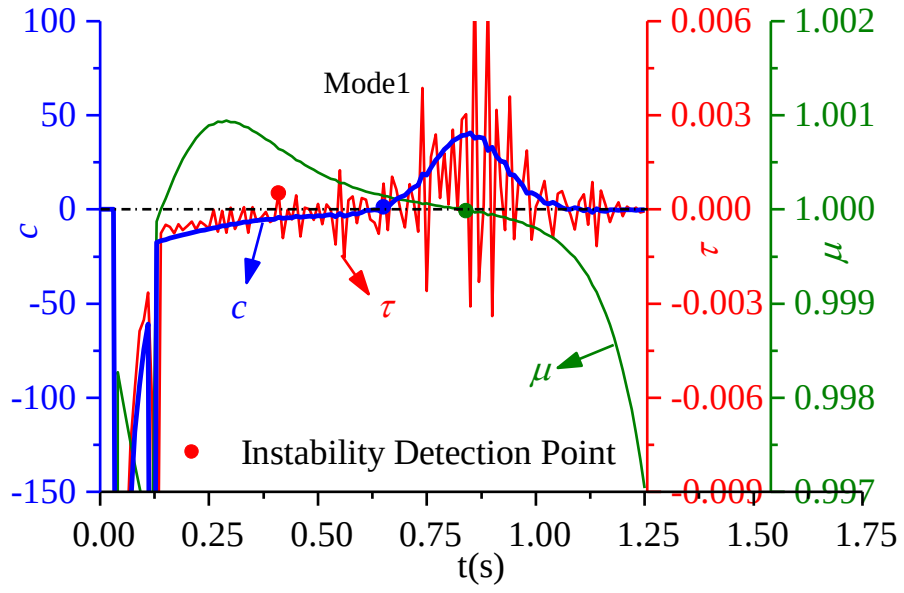

(b)

Fig. 7 Instability indexes comparison in (a) case 1 and (b) case 2

Fig. 7 (a) and (b) give the comparison of three instability indexes in case 1 and 2. For a fair comparison, the real-time

CMs identification method is also applied when we compute $\mu$ and $\tau$. Fig. 7 (a) shows that although the case 1 is stable, the index $\tau$ wrongly identifies the system as unstable because of the severe noise. And the misjudgments also occur to the index $\mu$. Only the proposed index $c$ can correctly identify the transient stability in case 1. Fig. 7 (b) shows all indexes can correctly identify the transient instability in case 2. However, the index $\mu$ takes more time than the proposed index $c$. And the severe noise of index $\tau$ significantly limits its application in practice. The comparison shows that the proposed instability index $c$ is superior to other instability indexes in identification accuracy and practicality.

To further verify the accuracy of the proposed scheme in general cases, we simulate three kinds of power flows, eight fault locations and two fault types in the 39-bus 10-machine power system. The faults are set at 0s, and then fault lines are cleared at 0.1s by the relays. Thus, we obtain 48 ($3\times8\times2$) cases, among which 32 cases are stable and 16 cases are unstable. A comparison of the proposed scheme and the emergency SIME method [3] is given in Table 3.

TABLE 3 COMPARISON OF EMERGENCY SIME METHOD AND PROPOSED CONCAVE-CONVEX AREA METHOD

| | Emergency SIME method | | | Concave-convex area method | | |
|---|---|---|---|---|---|---|
| Number of Cases | Identified as stable | Identified as unstable | Accuracy | Identified as stable | Identified as unstable | Accuracy |
| Stable cases(32) | 22 | 10 | 68.75% | 32 | 0 | 100% |
| Unstable cases(16) | 0 | 16 | 100% | 0 | 16 | 100% |

Table 3 shows that both the emergency SIME method and the proposed concave-convex area method can correctly identify the unstable cases. However, the emergency SIME method shows unsatisfactory accuracy in stable cases. The misjudgments are mainly caused by the insufficient prediction accuracy of the $\Delta P-\delta$ curve. In the emergency SIME method, a quadratic function is used to fit the $\Delta P-\delta$ curve. However, the practical $\Delta P-\delta$ curve is more complicated than the quadratic function due to the actions of regulators. The prediction errors may lead to inaccurate estimation of the maximum transient potential energy and finally result in the stability misjudgments. By contrast, the concave-convex area method is based on the real-time measurement data and does not require prediction. Thus, it shows higher accuracy and reliability.

## *B. 145 bus 50-machine power system*

The proposed instability index ensures the accurate start-up of real-time emergency control. However, the effectiveness of the proposed scheme still needs to be verified, that is, the start-up of emergency control is fast enough to restore the system to stability after the control. To this end, the started-up scheme is further tested in the 145-bus 50-machine power system. Fig. 8 shows this test system consists of 5 areas and 50 machines.

To examine the transient instability detection speed of the proposed start-up scheme, we test 20 unstable cases, and the results are given in Table 4. The table gives the average start-up angle $\delta_s$ and the average start-up time $t_s$ of the proposed scheme. As references, two threshold-based criteria are also given. Criterion 1 selects the $\delta_u$, and criterion 2 selects the fixed angle threshold of 180°. Criterion 1 is the symbol of transient instability in conventional TSA methods, and criterion 2 is the symbol of the out-of-step [29]. Table 4 shows that the proposed scheme can identify the transient instability 0.2024s before the criterion 1 and 0.4518s before the criterion 2. The fast instability detection of the proposed scheme leaves enough time for the emergency control implementation.

To verify the validity of the proposed start-up scheme, we simulate a complete scenario of real-time emergency control in case 3. The details of case 3 are given in Table 5 which contains the fault occurrence, fault clearance, emergency control start-up and implementation.

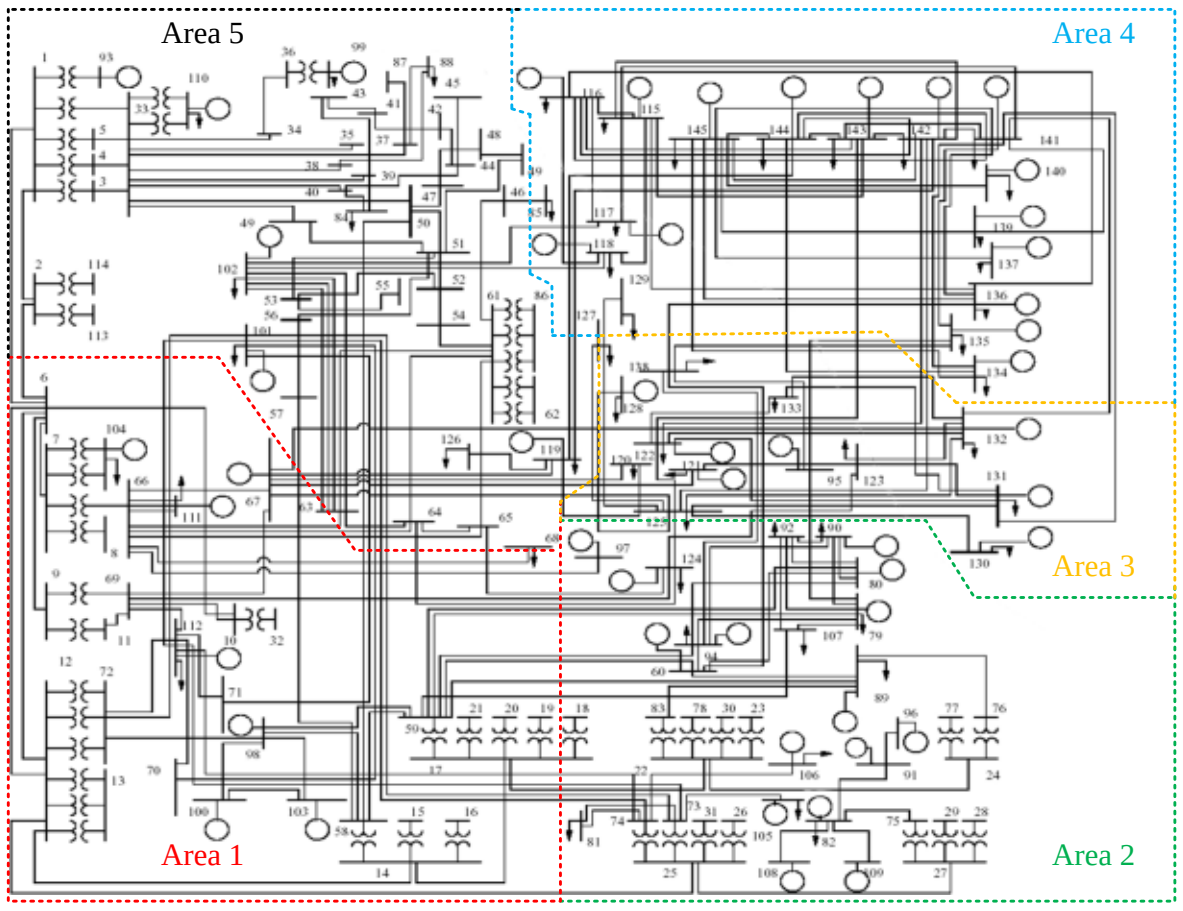

Fig. 8 145-bus 50-machine power system

TABLE 4 COMPARISON OF THE PROPOSED SCHEME AND OTHER CRITERIA ON AVERAGE START-UP ANGLE AND AVERAGE START-UP TIME

| | Proposed Start-up scheme | Criterion 1 | Criterion 2 |
|---|---|---|---|
| $\delta_s$ | 95.63° | 124.54° | 180° |
| $t_s$ | 0.3182s | 0.5206s | 0.7700s |
| Time ahead | - | 0.2024s | 0.4518s |

TABLE 5 DETAILS OF CASE 3

| Description | Case 3 |
|---|---|
| Fault details | A three-phase short-circuit fault occurs on line 105 at 0s; then the line is cleared at 0.1s. |
| Emergency control start-up time | 0.4s |
| CMs Identification | G42, G43 |
| Emergency Control Implementation time | 0.6s |
| Control Actions | G42 60%(1200MW) shedding |
| Final stability | Stable |

In case 3, the power system is unstable after the disturbance if no control is taken. The proposed scheme identifies the transient instability at 0.4s and then start the emergency control. The CMs identified at that moment are G42 and G43, based on which the control measures are determined using the trial and error method. A delay of 200ms is adopted in consideration of the communication delay and scheme computation time [36].

Thus, the control actions are implemented at 0.6s. After the emergency control, the power system returns to stability. The SMIB phase trajectories in Fig. 9 describe the entire transient process of case 3. Point *a*, *b*, *c*, and *d* represent the moments of fault occurrence, fault clearance, instability detection, and emergency control implementation, respectively. The trajectories in the red lines show the system becomes unstable without the emergency control. The blue trajectory indicates that the system returns to stability after the control. Case 3 illustrates that the proposed start-up scheme can initiate the emergency control fast enough to restore the system to stability.

Fig. 10 gives the comparison of different instability indexes in case 3. It shows that all indexes can detect the transient instability after the fault clearance. However, the index $\mu$ takes more time (0.56s) than the index $\tau$ (0.39s) and the proposed index *c* (0.4s). Besides, the system returns to stability after the emergency control, but both $\tau$ and $\mu$ wrongly indicate that the system is unstable after the control (See the instability detection points after time *d*). By contrast, the index *c* correctly indicates that the system is stable after the control. In short, the proposed instability index shows good accuracy and fast instability detection speed during the whole transient process.

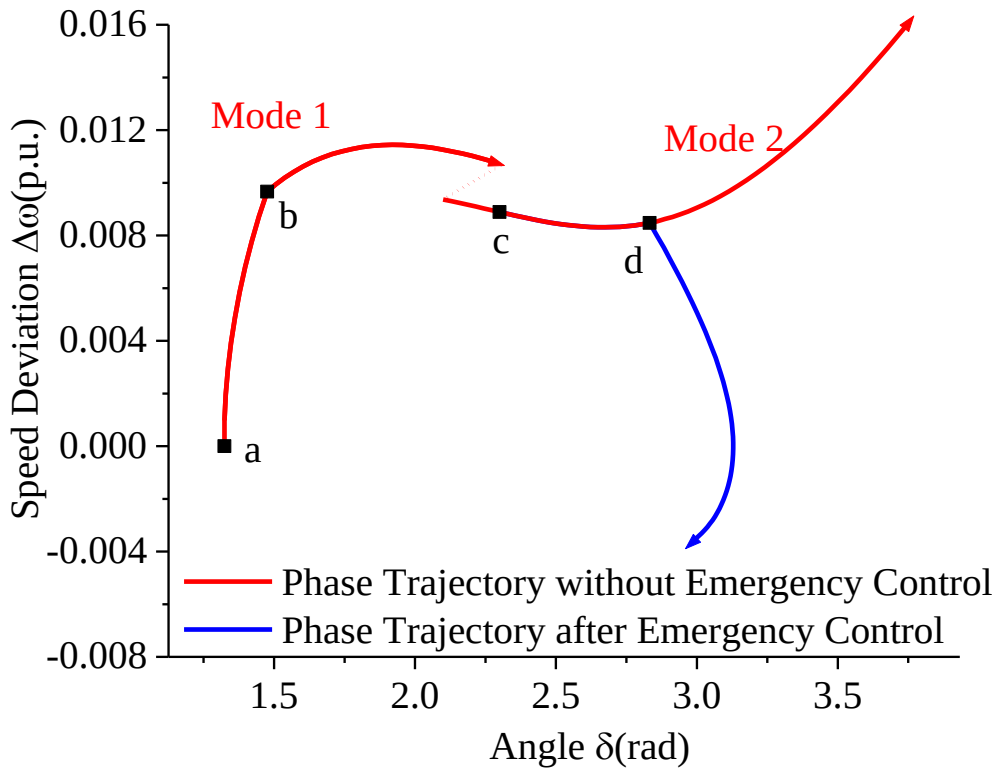

Fig. 9 The equivalent SMIB phase trajectories in case 3.
Point *a*: fault occurrence, *b*: fault clearance, *c*: instability detection and *d*: emergency control implementation

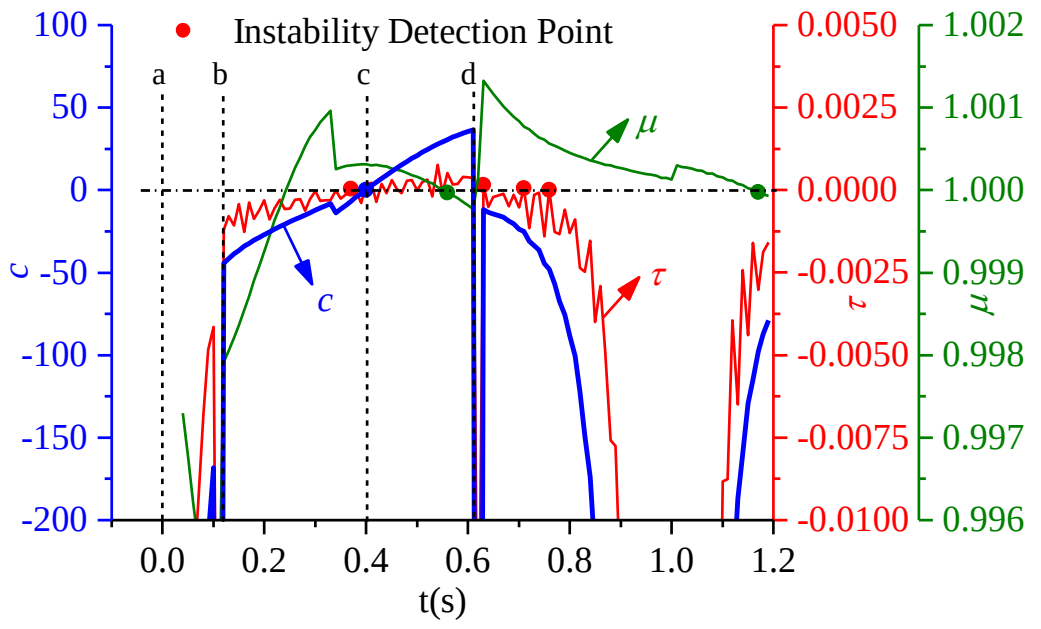

Fig. 10 Instability indexes comparison in case 3
Line *a*: fault occurrence, *b*: fault clearance, *c*: instability detection and *d*: emergency control implementation

### *C. Computation Efficiency Comparison*

Benefiting from novel two-layer SMIB equivalence framework, the proposed start-up scheme also shows high computation efficiency. In [36], the time for control start-up and decision is nearly 30ms. In [8], the offline and online computation time of the TSA method is 35.26ms and 7.6ms, respectively. By contrast, the average computation time of the proposed start-up scheme is only 3.227ms, as shown in Table 6. We obtain the total computation time by the sum of the maximum computation time in area control centers and the computation time in the global control center. All computation is done on MATLAB R2017a with an Intel i5 CPU. The computation time is expected to be even smaller with more efficient supercomputers.

TABLE 6 COMPUTATION TIME OF THE PROPOSED START-UP SCHEME

| Steps | Computation Time |
|---|---|
| **1st layer equivalence (maximum)** | |
| Area generators grouping | 0.47ms |
| COI equivalence | 1.20ms |
| **2nd layer equivalence** | |
| CMs identification | 0.43ms |
| SMIB equivalence | 1.10ms |
| **Index computation and judgement** | 0.027ms |
| Total | 3.227ms |

With the increase in the number of generators, the advantage of high computation efficiency of the proposed two-layer SMIB equivalence framework becomes more prominent. Fig. 11 gives the computation time comparison between the direct SMIB equivalence method and the proposed two-layer SMIB equivalence method. The figure shows that the computation time of the direct equivalence method, $T_{Direct}$, grows rapidly with the increase in generator numbers. By contrast, the computation of the proposed two-layer equivalence method still only takes a small amount of time. For example, $T_{Direct}$ increases to 36.1ms when the number of generators is 2000. However, the computation time using the two-layer SMIB equivalence method, $T_{Two\text{-}layer}$, is only 4.9ms. The computation efficiency of the two-layer SMIB equivalence method is 7.37 times that of direct SMIB equivalence method. In the proposed scheme, generators are separated into 20 areas, and each area has an average of 100 generators. The total computation time is the sum of the maximum computation time of 1st layer equivalence (100 generators) in the area control centers and the time of 2nd layer equivalence (20-40 COI generators) in the global control center. The area control centers share most of the computational tasks through parallel computing, which is the key to the high efficiency of the proposed method.

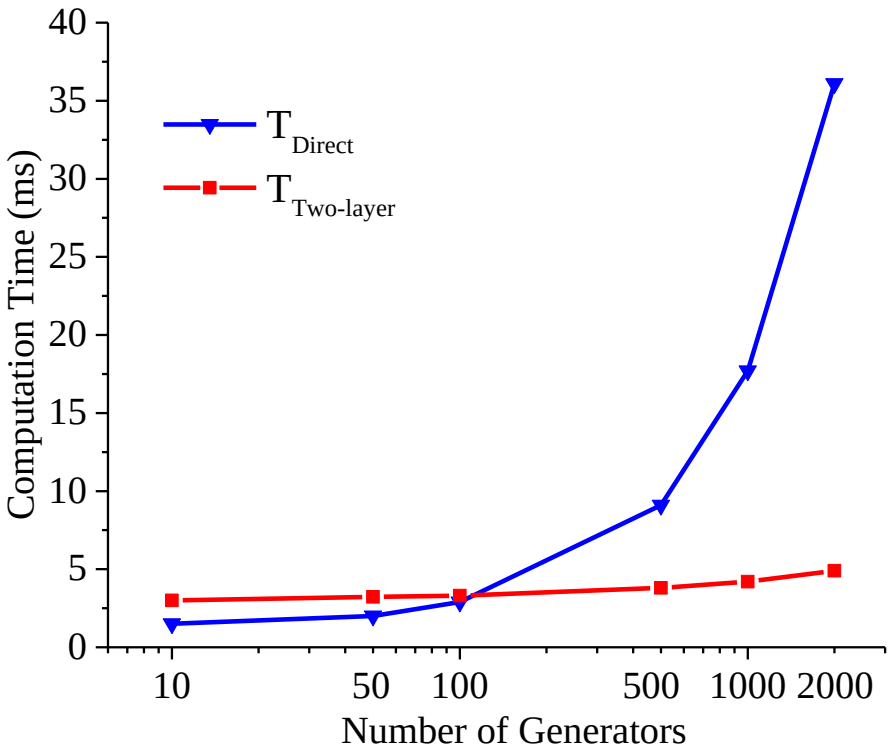

Fig. 11 Computation time comparison between the direct SMIB equivalence method and the proposed two-layer SMIB equivalence method

In addition, the two-layer SMIB equivalence method also significantly reduces the communication burden. For the 2000-machine power system, all generator data needs to be uploaded to the global control center in the conventional direct SMIB equivalence method. However, in the proposed two-layer SMIB equivalence method, only the data of 20-40 COI generators needs to be transmitted to the global control center. The volume of the data to be transmitted in the new equivalence method is only 1%-2% of the one in the conventional equivalence method. The reduction in communication burden will improve the efficiency and accuracy of data transmission, which undoubtedly benefits the fast and accurate start-up of real-time emergency control.

### *D. Performance under Measurement Errors*

Above simulations assume that all PMU measurement data is accurate. However, the measurement errors are unavoidable in the practical power system. It is thus worthwhile to assess the performance of the proposed method in the presence of measurement errors.

In this paper, the Gaussian white noise is used as the PMU measurement errors. The noise is directly added to the original measurement data including $\delta, \Delta\omega$ and $\Delta P$. The IEEE standard requires that the total vector error (TVE) of PMU measurement data should be less than 1% [31]. According to the definition of signal-to-noise ratio (SNR) in (12), the commercial standard for PMU measurements is SNR>40dB.

$$SNR_{dB} = 20\lg\left(\frac{A_{signal}}{A_{noise}}\right) \quad (12),$$

where $A_{signal}$ and $A_{noise}$ are the amplitude of signal and noise, respectively.

To fully verify the effectiveness of the proposed method under measurement errors, we test it under different levels of errors (40dB, 50dB, and 100dB), and find that the proposed instability index shows high noise tolerance. Case 4 is given as an example to show the performance of the proposed instability index under measurement errors.

Case 4: In the 39-bus 10-machine power system, a three-phase short-circuit ground fault occurs on the line 16-17 at $t$=0s, then the fault line is eliminated at 0.1s. The system is stable after the disturbance.

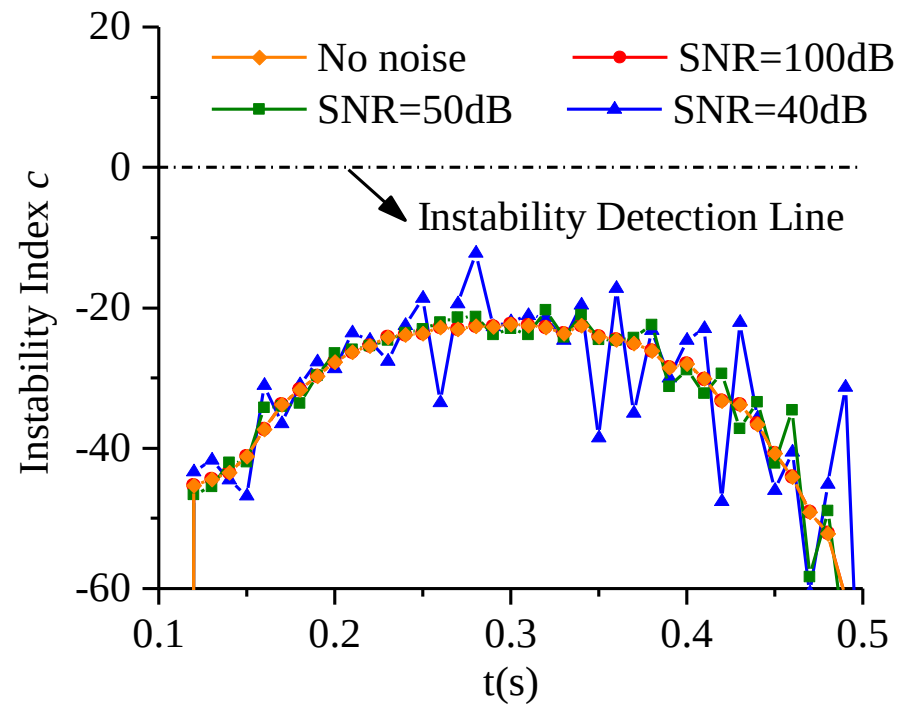

Fig. 12 Performance of the proposed instability index under different levels of measurement errors

The instability index curves under different levels of errors are given in Fig. 12. It shows that the proposed method, without any noise reduction measures, has nearly the same accuracy in scenarios where SNR=100dB as in scenarios without noise. In scenarios with more serious errors where SNR=50dB or SNR=40dB, the instability index can still correctly indicate the transient stability of the power system. By contrast, other instability indexes, $\tau$ and $\mu$, encounter serious misjudgments under measurement errors (SNR=40dB), as shown in Fig. 13. In short, the proposed method is robust to noise and is qualified to be a start-up scheme for real-time emergency control in the presence of allowable PMU measurement errors.

However, it should be noted in Fig. 12 that as the error level increases, the fluctuations of the instability index curves become apparent. This phenomenon increases the risks of misjudgments in stable scenarios. For example, when the error level increases to SNR=30dB, TVE=3.162%, the system is wrongly detected as unstable in case 4, as shown in Fig. 14. To address the misjudgments caused by severe measurement errors, two noise reduction measures are suggested in this paper: 1) use moving average filter (MAF) on the original measurement data and 2) use MAF on the instability index curve. The results are given in Fig. 14. The figure indicates that either of the noise reduction measures can avoid the misjudgments in scenarios with severe measurement errors.

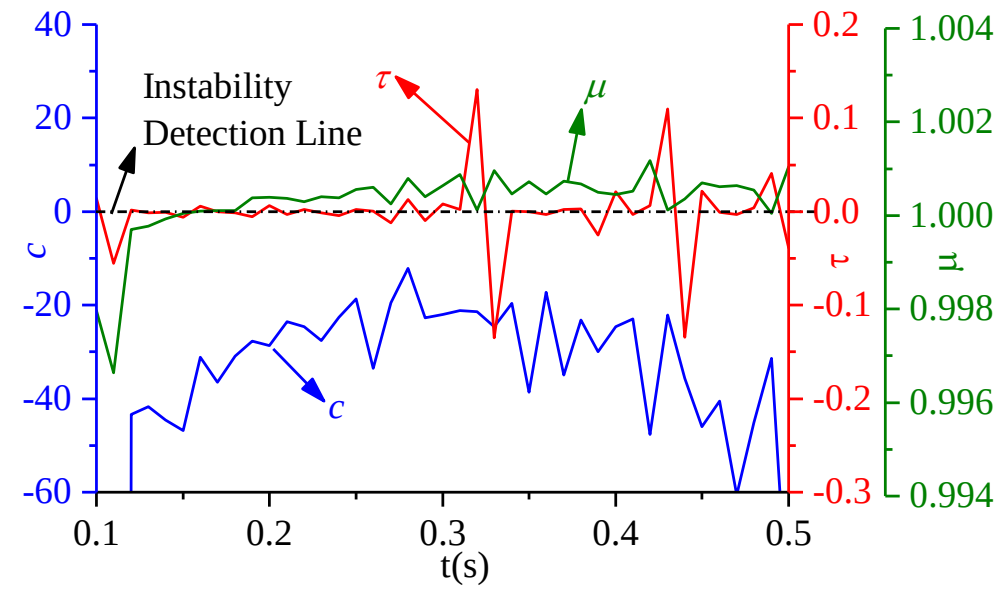

Fig. 13 Instability Index Comparison under measurement errors (SNR=40dB)

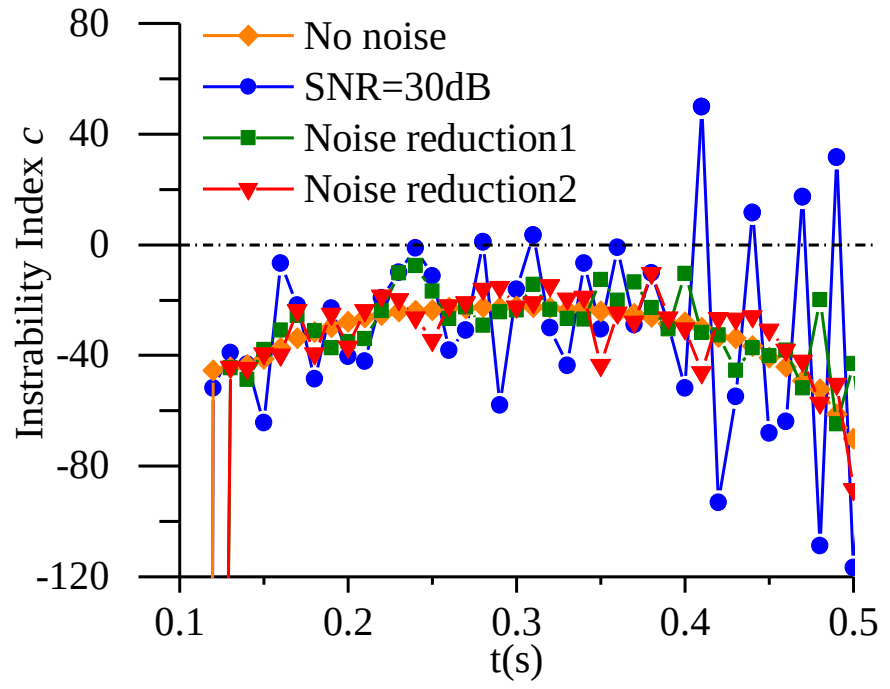

Fig. 14 Performance of the proposed instability index under severe errors (SNR=30dB, TVE=3.162%) and noise reduction measures

## V. CONCLUSIONS

In this paper, an accurate and fast start-up scheme is proposed for the real-time emergency control. This work aims to identify the transient instability of the power system with the real-time measurement data from PMUs and start the emergency control to restore the system to stability.

To achieve this goal, a new instability index is firstly proposed based on the concave-convex area. It shows superiority in accuracy, reliability and noise tolerance when compared with other instability indexes. Besides, a real-time CMs identification algorithm is used to track the changes of CMs. It ensures the accuracy of SMIB equivalence at each moment. Then a novel two-layer SMIB equivalence framework is presented to improve the computation efficiency and reduce the communication burden. The framework greatly promotes the real-time application of the proposed method in the large-scale power system. Finally, the simulations in two test systems verify the effectiveness of the scheme.

## REFERENCE

[1] J. J. Q. Yu, D. J. Hill, A. Y. S. Lam, J. Gu and V. O. K. Li, "Intelligent Time-Adaptive Transient Stability Assessment System," *IEEE Trans Power Syst.*, vol.33, no.1, pp.1049-1058, 2018

[2] M. A. M. Ariff and B. C. Pal, "Adaptive Protection and Control in the Power System for Wide-Area Blackout Prevention," *IEEE Trans. Power Delivery*, vol.31, no.4, pp.1815-1825, 2016

[3] M. Pavella, D. Ernst and D. Ruiz-Vega, *Transient stability of power systems: a unified approach to assessment and control*, Berlin, Germany: Springer Science & Business Media, 2012,

[4] M. Oluic, M. Ghandhari and B. Berggren, "Methodology for Rotor Angle Transient Stability Assessment in Parameter Space," *IEEE Trans. Power Syst.*, Vol.32, No.2, pp. 1202-1211, 2017

[5] J. Gou, Y. Liu, J. Liu, G. A. Taylor and M. M. Alamuti, "Novel pair-wise relative energy function for transient stability analysis and real-time emergency control," *IET Gener., Transm. Distrib.*, Vol.11, No.18, pp. 4565-4575, 2017

[6] X. Xu, H. Zhang, C. Li, Y. Liu, W. Li and V. Terzija, "Optimization of the Event-Driven Emergency Load-Shedding Considering Transient Security and Stability Constraints," *IEEE Trans. Power Syst.*, vol.32, no.4, pp.2581-2592, 2017

[7] Z. Li, G. Yao, G. Geng and Q. Jiang, "An Efficient Optimal Control Method for Open-Loop Transient Stability Emergency Control," *IEEE Trans. Power Syst.*, vol.32, no.4, pp.2704-2713, 2017

[8] P. Bhui and N. Senroy, "Real Time Prediction and Control of Transient Stability Using Transient Energy Function," *IEEE Trans. Power Syst.*, Vol.32, No.2, pp. 923-934, 2017

[9] S. Fang, H. Zhang and G. Xue, "Instability prediction of the inter-connected power grids based on rotor angle measurement," *Int. J. Elec. Power Energ. Syst.* , vol.88, pp.21-32, 2017

[10] D. Huang, X. Yang, S. Chen, and T. Meng, "Wide-area measurement system-based model-free approach of post-fault rotor angle trajectory prediction for on-line transient instability detection", *IET Gener., Transm. Distrib.*, vol. 12, no. 10, pp. 2425-2435, 2018.

[11] A. Shamisa, B. Majidi, and J. Patra, "Sliding-Window-Based Real-Time Model Order Reduction for Stability Prediction in Smart Grid", *IEEE Trans. Power Syst.*, pp. 1, 2018.

[12] G. Qun and S. M. Rovnyak, "Decision Trees Using Synchronized Phasor Measurements for Wide-Area Response-Based Control," *IEEE Trans. Power Syst.*, vol.26, no.2, pp.855-861, 2011

[13] M. He, V. Vittal and J. Zhang, "Online dynamic security assessment with missing pmu measurements: A data mining approach," *IEEE Trans. Power Syst.*, vol.28, no.2, pp.1969-1977, 2013

[14] A. G. Bahbah and A. A. Girgis, "New Method for Generators' Angles and Angular Velocities Prediction for Transient Stability Assessment of Multimachine Power Systems Using Recurrent Artificial Neural Network," *IEEE Trans. Power Syst.*, vol.19, no.2, pp.1015-1022, 2004

[15] R. Zhang, Y. Xu, Z. Dong and K. P. Wong, "Post-disturbance transient stability assessment of power systems by a self-adaptive intelligent system," *IET Gener. Transm. Distrib.*, vol.9, no.3, pp.296-305, 2015

[16] S. Mehraeen, S. Jagannathan and M. L. Crow, "Power System Stabilization Using Adaptive Neural Network-Based Dynamic Surface Control," *IEEE Trans. Power Syst.*, vol.26, no.2, pp.669-680, 2011

[17] AD. Rajapakse, F. Gomez, K. Nanayakkara, P. A. Crossley and V. V. Terzija, "Rotor Angle Instability Prediction Using Post-Disturbance Voltage Trajectories," *IEEE Trans. Power Syst.*, vol.25, no.2, pp.947-956, 2010

[18] F. R. Gomez, AD. Rajapakse, U. D. Annakkage and IT. Fernando, "Support Vector Machine-Based Algorithm for Post-Fault Transient Stability Status Prediction Using Synchronized Measurements," *IEEE Trans. Power Syst.*, vol.26, no.3, pp.1474-1483, 2011

[19] B. Wang, B. Fang, Y. Wang, H. Liu and Y. Liu, "Power System Transient Stability Assessment Based on Big Data and the Core Vector Machine," *IEEE Trans. Smart Grid*, vol.7, no.5, pp.2561-2570, 2016

[20] J. C. Cepeda, J. L. Rueda, D. G. Colome and D. E. Echeverria, "Real-time transient stability assessment based on centre-of-inertia estimation from phasor measurement unit records," *IET Gener. Transm. Distrib.*, vol.8, no.8, pp.1363-1376, 2014

[21] S. Dasgupta, M. Paramasivam, U. Vaidya and V. Ajjarapu, "PMU-Based Model-Free Approach for Real-Time Rotor Angle Monitoring," *IEEE Trans. Power Syst.*, vol.30, no.5, pp.2818-2819, 2015

[22] D. R. Gurusinghe and A. D. Rajapakse, "Post-Disturbance Transient Stability Status Prediction Using Synchrophasor Measurements," *IEEE Trans. Power Syst.*, vol.31, no.5, pp.3656-3664, 2016

[23] A. Shamisa and M. Karrari, "Model free graphical index for transient stability limit based on on-line single machine equivalent system identification," *IET Gener. Transm. Distrib.*, vol.11, no.2, pp.314-321, 2017

[24] L. Wang and A. A. Girgis, "A new method for power system transient instability detection," *IEEE Trans. Power Delivery*, vol.12, no.3, pp.1082-1089, 1997

[25] H. Xie, B. Zhang, G. Yu, Y. Li, P. Li, D. Zhou and F. Yao, "Power System Transient Stability Detection Based on Characteristic Concave or Convex of Trajectory," in *Transmission and Distribution Conference and Exhibition: Asia and Pacific, 2005 IEEE/PES*, 2005, PP. 1-6

[26] B. Zhang, S. Yang, H. Wang, S. Ma and L. Wu, " Closed-loop control of power system transient stability (2): transient instability detection method of multi-machine power system," *Electr. Power Autom. Equip.*, no.09, pp.1-6, 2014 (in Chinese)

[27] Y. Xu, Z. Yang, J. Zhang, Z. Fei, and W. Liu, "Real-Time Compressive Sensing based Control Strategy for a Multi-area Power System", *IEEE Trans. Smart Grid*, vol.9 no.5, pp. 4293-4302, 2018

[28] S. Zhao, H. Jia, and D. Fang, "Partition-composition method for online detection of interconnected power system transient stability", *IET Gener. Transm. Distrib.*, vol. 10, no. 14, pp. 3529-3538, 2016.

[29] P. Kundur, N. J. Balu and M. G. Lauby, *Power system stability and control*, New York: McGraw-hill, 1994,

[30] D. E. Echeverria, J. C. Cepeda and D. G. Colome, "Critical machine identification for power systems transient stability problems using data mining," in *IEEE PES Transmission & Distribution Conference and Exposition-Latin America (PES T&D-LA), 2014*, 2014,

[31] IEEE Standard for Synchrophasor Measurements for Power Systems, C37.118.1, 2011

[32] J. Machowski, J. Bialek, J.R. Bumby, and J. Bumby, *Power system dynamics and stability*, John Wiley & Sons, 1997.

[33] IEEE 39-bus 10-machine power system model. [Online]. Available: http://www.sel.eesc.usp.br/ieee/index.htm

[34] V. Vittal, D. Martin, R. Chu, J. Fish, J. C. Giri, C. K. Tang, F. Eugenio Villaseca and R. G. Farmer, "Transient stability test systems for direct stability methods," *IEEE Trans. Power Syst.*, vol.7, no.1, pp.37, 1992

[35] IEEE recommended practice for excitation system models for power system stability studies." IEEE Std 421.5™, 2016.

[36] W. Yu, Y. Xue, J. Luo, M. Ni, H. Tong and T. Huang, "An UHV Grid Security and Stability Defense System: Considering the Risk of Power System Communication," IEEE Trans. Smart Grid, vol.7, no.1, pp.491-500, 2016

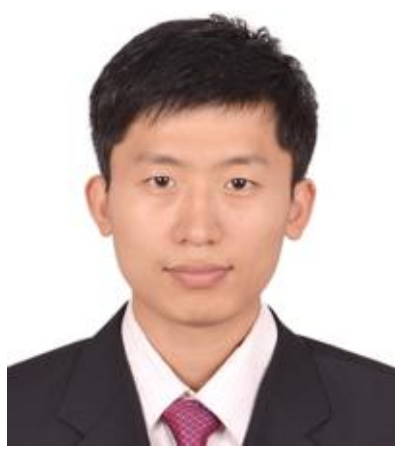

**Songhao Yang** (S'18) received the B.S. degree in electrical engineering from Xi'an Jiaotong University, Xi'an, China, in 2012. He is currently working toward the Ph.D. degree at the School of Electrical Engineering, Xi'an Jiaotong University. Meanwhile, he is also pursuing his Ph. D. degree in engineering at Tokushima University.

His main field of interests include optimal PMUs placement, power system stability assessment and control.

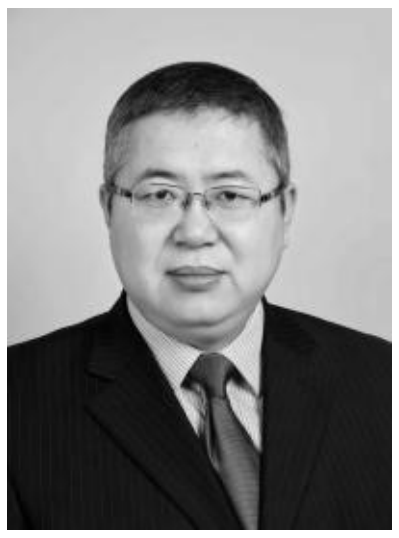

**Zhiguo Hao** (M'10) was born in Ordos, China, in 1976. He received the B.Sc. and Ph.D. degrees in electrical engineering from Xi'an Jiaotong University, Xi'an, China, in 1998 and 2007, respectively.

He has been an Associate Professor with the Electrical Engineering Department, Xi'an Jiaotong University, since 2013. His research interest includes power system protection.

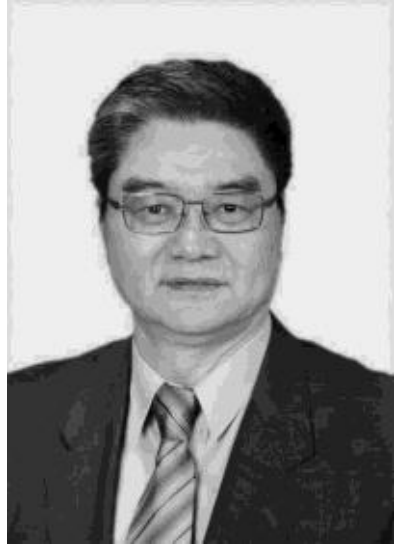

**Baohui Zhang** (SM'99-F'19) was born in Hebei, China, in 1953. He received the M. Eng. and Ph.D. degrees in electrical engineering from Xi'an Jiaotong University, Xi'an, China, in 1982 and 1988, respectively.

Since 1992, he has been a Professor with the Electrical Engineering Department, Xi'an Jiaotong University. His research interests include power system analysis, control, communication, and protection.

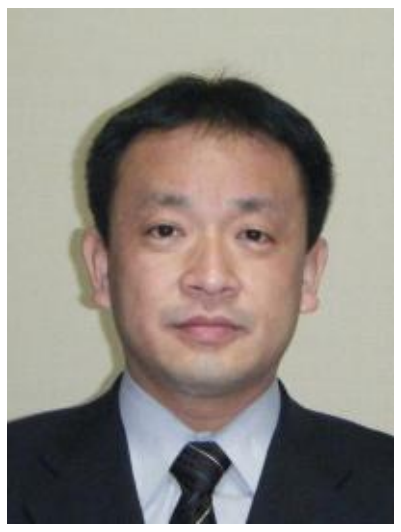

**Masahide Hojo** (S'98-M'99) was born Tokushima, Japan. He received the M.S. degree in engineering from Kobe University in 1996 and the Ph.D. degree in engineering from Osaka University in 1999.

He is currently a professor at Tokushima University. His research interests include the advanced power system control by power electronics technologies, and analysis of power systems.